\documentclass{ws-procs975x65} 
\newcommand{\be}{\begin{equation}}
\newcommand{\ee}{\end{equation}}
\newcommand{\ba}{\begin{eqnarray}}
\newcommand{\ea}{\end{eqnarray}}

\makeatletter
\newcommand\xleftrightarrow[2][]{%
  \ext@arrow 9999{\longleftrightarrowfill@}{#1}{#2}}
\newcommand\longleftrightarrowfill@{%
  \arrowfill@\leftarrow\relbar\rightarrow}
\makeatother

\begin{document}
\title{Topology, the meson spectrum and the scalar glueball: \\ 
three probes of conformality and the way it is lost. }

\author{E. Pallante$^*$}

\address{Van Swinderen Institute, University of Groningen,\\
Groningen, 9747 AG, The Netherlands \\
$^*$E-mail: e.pallante@rug.nl }


\begin{abstract}
We discuss properties of non-Abelian gauge theories that change significantly across the lower edge of the conformal window. Their probes are the topological observables, the meson spectrum and the scalar glueball operator. The way these quantities change tells about the way conformal symmetry is lost. 
\end{abstract}


\bodymatter

\section{Introduction}\label{sec:intro}

It is a compelling task to understand the role played by conformal symmetry in the yet to be discovered unified  theory for gravity, electroweak and strong interactions. 
Instrumental to this aim and relevant for phenomenological constructions beyond the Standard Model is the study of the  breaking patterns of this symmetry in each component of the complete theory taken in isolation as well as when coupled to each other. 

In this work we mainly focus on the conformal window of non-Abelian gauge theories in isolation, and in order to simplify the discussion we consider the case of $SU(N)$ Yang-Mills theory with massless Dirac fermions in the fundamental representation, while explicitly commenting on other realizations whenever useful. Towards the end, in Sec.~\ref{sec:spectrum}, we come back to some considerations about the embedding in the complete theory.  

\begin{figure}[h]
\begin{center}
\includegraphics[width=3in]{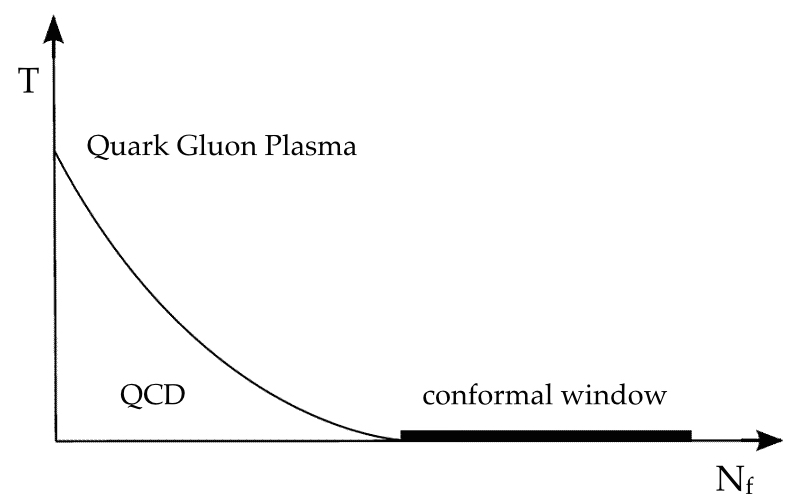}
\end{center}
\caption{Temperature (T)\,--\,flavor ($N_f$) phase diagram of $SU(N)$ Yang-Mills theory with $N_f$ massless Dirac fermions in the fundamental representation. A line of thermal chiral phase transitions, the chiral phase boundary, separates QCD from the high-temperature quark-gluon plasma phase. The endpoint of the chiral phase boundary is also where the theory deconfines and it coincides with the lower edge $N_f^c$ of the conformal window; the addition of supersymmetry and/or different fermion representations can change the latter property and make the phase diagram richer. }
\label{fig:phase}
\end{figure}

The phase diagram of massless QCD in the plane temperature versus the number of flavors is shown in Fig.~\ref{fig:phase}. We are interested in the change of properties of the system across the lower edge of the conformal window; the latter defines a family of theories that has a stable infrared fixed point (IRFP) where the theory is conformal with nonzero anomalous dimensions. Theories inside the conformal window are deconfined and chiral symmetry is exact. 
In Sec.~\ref{sec:glueball} we discuss properties of the scalar glueball operator and how its anomalous dimension probes different mechanisms for the closing of the conformal window. Sec.~\ref{sec:topology} discusses the topology, the eigenvalue distribution of the Dirac operator and the role of the $U(1)$ axial anomaly. In Sec.~\ref{sec:spectrum} we entertain the possibility of a spontaneous breaking of conformal symmetry in the complete theory and discuss under which conditions a dilaton could manifest in the QCD sector.

\section{The scalar glueball and its anomalous dimension}
\label{sec:glueball}

The lower edge of the conformal window of QCD as depicted in Fig.~\ref{fig:phase} can be seen  
as the point in the space of relevant couplings where the system at the same time deconfines and realizes exact chiral symmetry at zero temperature. There is no physical reason, nor evidence for a separation of where the system deconfines and where chiral symmetry is restored; if it were not so, a new phase would appear below the conformal window, separated from QCD by an additional change of symmetry patterns\footnote{Differently from QCD, supersymmetric QCD (SQCD) develops a free non-Abelian magnetic phase in terms of the dual magnetic variables for $N_c+2\leq N_f\leq 3N_c/2$, just below the conformal window where the fundamental electric theory is infinitely strongly coupled. 
 This can be seen as a consequence of exact electric-magnetic duality\cite{Seiberg:1994pq}.}. 
In fact, the recent study in Ref.~\citenum{daSilva:2015vna} provides close
 agreement between the lattice determination of the lower edge of the conformal window using chiral observables $6<N_f^c<8$ and the prediction in 
Ref.~\citenum{Bochicchio:2013aha}, where  large-$N$ QCD in the Veneziano limit ($N\rightarrow\infty$, $N_f/N={\mbox{const}}$)  is solved and the lower edge is identified as the threshold for the quantum instability of the glueball kinetic term $N_f/N=5/2$.

The properties of gluodynamics must change across the lower edge of the conformal window, where conformal symmetry is restored and the system realizes an exact IRFP. 
In particular, the gluon condensate, i.e., the vacuum expectation value of the scalar glueball operator $ G^a_{\mu\nu}G^a_{\mu\nu}$ must identically vanish at the fixed point and the two-point correlator of the scalar glueball acquires conformal scaling with  nonzero anomalous dimension $\gamma_G$. 
In Ref.~\citenum{daSilva:2015vna} we realized that $\gamma_G$, for varying $N_f$ along the IRFP line, carries information on the nature of the endpoint and the 
zero-temperature phase that precedes the conformal window. 
At a fixed point the beta-function vanishes $\beta (\alpha_*)=0$ and one finds \cite{daSilva:2015vna}
\be
\gamma_G= -\beta' (\alpha_*)\, .
\ee
The latter  is a physical property of the system, renormalization scheme independent. 
In Ref.~\citenum{daSilva:2015vna} we have determined 
 $\gamma_G$ in perturbation theory to a given loop order, $n=2,\,3,\,4$ loops, inside the conformal window and compared this result with the Veneziano limit of 
large-$N$ QCD\cite{Bochicchio:2013aha}. 
One salient aspect is the fact that 
perturbation theory  as well as the large-$N$ solution  predict an {\em increasing} anomalous dimension $\vert \gamma_G\vert$ $(\vert\beta'(\alpha_*\vert )$ for {\em decreasing} $N_f{\small{\searrow}}N_f^c$. 
This behavior turns out to be  opposite\cite{daSilva:2015vna} to the one implied by a UV-IR fixed-point merging phenomenon at $N_f^c$. In this case we can assume  without loss of generality 
\be
\label{eq:UVIF}
\beta(\alpha, N_f) = (N_f-N_f^c) - (\alpha -\alpha^c)^2
\ee
close to $N_f^c$ where the merging would occur, see also Ref.~\citenum{Kaplan:2009kr}. This beta-function has pairs of zeros 
$\alpha_{\pm} = \alpha^c\pm\sqrt{N_f-N_f^c}$ and it
develops a local maximum $\beta'(\alpha^c)=0$   for $N_f=N_f^c$,  with $\vert\beta'\vert$  {\em decreasing}  for {\em decreasing} $N_f{\small{\searrow}}N_f^c$.  Both zeros $\alpha_{\pm}$ acquire an imaginary component for $N_f<N_f^c$. 

A UV-IR merging mechanism is the simplest, maybe not unique, framework where the so called walking phenomenon is realized just below the conformal window. Instead of using the running coupling\,---\,a renormalisation scheme dependent quantity\,---\,this phenomenon is best identified by the realization of preconformal Miransky/Berezinsky-Kosterlitz-Thouless (BKT) scaling\cite{Miransky:1984ef, Miransky:1996pd, Kosterlitz:1974sm} of observables  in the vicinity of the conformal window\footnote{The comparison of theories with different $N_f$ requires the setting of a common reference scale. A common UV scale is to be chosen in order to trace a preconformal Miransky/BKT scaling in the IR-dominated observables.}, for $N_f\nearrow N_f^c$.
From a statistical mechanics point of view, the realization of this scaling is a direct consequence of a so-called conformal phase transition that would occur at  $N_f^c$, treating $N_f$ as a continuous parameter. The inverse of the correlation length $\xi$ of the system 
\be
\frac{1}{\xi} \sim \Lambda_{UV}\,\theta (N_f^c-N_f)\, e^{-{c}/{\sqrt{N_f^c-N_f}}}\, ,~~~~~~c>0
\ee
decreases exponentially in the  broken phase $N_f<N_f^c$ and it vanishes identically in the symmetric phase $N_f>N_f^c$, leaving an imprint in the spectrum on both sides. We come back to this point in 
Sec.~\ref{sec:spectrum} when studying the spectrum across the endpoint and the role of conformal symmetry. 

A scenario where no UV-IR fixed point merging occurs remains a plausible one.  It is actually favored by perturbation theory and it more closely resembles what happens in SQCD, where no physical UVFP 
emerges\,---\,differences can still reside in the detailed structure of the beta-function, as noticed in 
Ref.~\citenum{Bochicchio:2013aha} and further discussed in Ref.~\citenum{daSilva:2015vna}, and in the occurrence of the free magnetic phase in SQCD. 

A nonperturbative, e.g., lattice determination of the anomalous dimension $\gamma_G$ is therefore desirable, in order to discriminate between realizations of loss  of conformality with and without UV-IR merging. 
One obvious way to extract $\gamma_G$ is by a nonperturbative measurement of the two-point correlator of the scalar glueball operator; on the lattice one should separate its universal scaling law from nonuniversal violations of scaling.  
Another intriguing possibility may be suggested by the study of the Wilson flow of the vacuum expectation value of the scalar glueball  operator (the gauge energy density) in Ref.~\citenum{Luscher:2010iy}.  
There, the evolution of this expectation value away from the ultraviolet has been derived in perturbative QCD as the solution of a Wilson flow equation, an approximately trivializing map realized by a first order ordinary differential equation in field space with the Wilson gauge action as the evolution kernel. Trivializing maps have been discussed in the context of field theories in the continuum and on the lattice\cite{Narayanan:2006rf}, while Ref.~\citenum{Luscher:2010iy} discusses explicit examples and proposes their use for high precision scale setting in lattice QCD studies, see also Refs.~\citenum{Luscher:2011kk, Borsanyi:2012zs}. 
By defining the quantity $t^2\langle E\rangle$, with $t$ the flow ``time" with dimensions $\left [{\mbox{energy}}^{-2}\right ]$, in the $\overline{MS}$ scheme and at a renormalization scale $\mu= 1/\sqrt{8t}$ one obtains for $SU(3)$\cite{Luscher:2010iy}
\ba
\label{eq:pertQCD}
&&\hspace{-0.4cm}t^2 \langle E\rangle =\frac{3}{4\pi}\alpha(\mu)\left\{ 1+k_1\alpha (\mu) +O(\alpha^2)\right\}\nonumber\\
&&\hspace{-0.4cm}k_1= \frac{1}{4\pi}\left\{ 11\gamma_E +\frac{52}{3}-9\ln{3} - N_f \left ( \frac{2}{3}\gamma_E +\frac{4}{9}-\frac{4}{3}\ln{2}\right )\right\} \nonumber\\
&&\hspace{-0.4cm}\phantom{k_1}= 1.0978 +0.0075\,N_f
\ea
to next-to-leading order in the  running coupling $\alpha (\mu)=g^2(\mu)/(4\pi )$. The quantity in Eq.~(\ref{eq:pertQCD}) exhibits the important property of being finite once expressed in terms of the renormalized gauge coupling, at least at next-to-leading order in perturbation theory.  
In other words, the flow of  $t^2\langle E\rangle\rightarrow 0$  on a Euclidean  spacetime lattice as a function of $t$ is dictated by dimensional arguments, while its proportionality to the running coupling in perturbation theory encodes asymptotic freedom. 
In particular, the residual $t$-scale dependence in the running coupling encodes the specific and highly nontrivial way in which asymptotic freedom realizes the breaking of conformal symmetry in the quantized theory; asymptotic freedom also guarantees that $t^2\langle E\rangle\rightarrow 0$ in the ultraviolet limit $t\rightarrow 0$. 
Beyond perturbation theory, we could think to infer the evolution towards the infrared of a properly renormalized $t^2\langle E\rangle$ based, again, on dimensional reasoning and underlying symmetries. However, constraints on one-point functions are usually not as powerful as those on other $n-$point functions. To illustrate this we use an argument that we argue should be valid for both $\bar{\psi}\psi$ and $GG$ at an IRFP. Both VEVs $\langle\bar\psi\psi\rangle$ and $\langle GG\rangle$ must vanish at the IRFP.  
One would like to know  how they approach zero when a scale perturbation\,---\,it being the fermion mass or $t$, respectively\,---\,is applied.  Chiral symmetry is exact in the conformal window, and only at the lower endpoint of the IRFP line a chiral restoration phase transition occurs. In other words, the chiral order parameter $\langle \bar\psi\psi\rangle$ along this line, except its endpoint, does not undergo a phase transition and it must obey all constraints of exact chiral symmetry. One constraint  that will also be later useful is the one on the chiral cumulant $R_\pi=\chi_\sigma/\chi_\pi=(\partial\langle \bar\psi\psi\rangle /\partial m)/ (\langle \bar\psi\psi\rangle /m)$, or equivalently the ratio of the renormalized  pseudoscalar and scalar meson masses. At the chiral restoration transition the relevant magnetic EoS\cite{Deuzeman2010} 
\begin{equation}
\label{eq:EoS}
m=A\langle \bar\psi\psi\rangle^\delta +B \langle \bar\psi\psi\rangle + \langle \bar\psi\psi\rangle_0
\end{equation}
has $B=0$, $\langle \bar\psi\psi\rangle_0 =0$ and it implies $R_\pi=1/\delta$, with critical exponent $\delta$. Chiral symmetry implies that $\lim_{m\to 0} R_\pi =0$ in the broken phase, while  $\lim_{m\to 0} R_\pi =1$ in the symmetric phase; thus $0\leq R_\pi\leq 1$. Away from the endpoint along the IRFP line no chiral phase transition occurs, and 
exact chiral symmetry suggests  $\lim_{m\to 0} R_\pi =1$; thus $B\neq 0$. Differently from the endpoint, the only constraint imposed by conformal symmetry on the scaling 
of this and other one-point functions
with a scale perturbation  is that their VEVs vanish at the IRFP;  this is guaranteed for $\langle \bar\psi\psi\rangle$ also with $B\neq 0$. 
On the other hand, given that $0\leq 1/\delta \leq 1$, the contribution of the $A$ term is dominant w.r.t. the linear term close to the chiral limit. An analogous reasoning can be applied to $\langle GG\rangle$ ($\langle E\rangle$), where now chiral symmetry is no longer imposing constraints.  The $t$-evolution can contain a term carrying its anomalous dimension $t^{-2-\gamma_G/2}$, but it is not the only one. Also note that this term goes to zero for $t\rightarrow\infty$ only if $\gamma_G>-4$. 

Obviously, the infrared behavior realized on the lattice by the quantity $t^2\langle E\rangle$ will depend on the chosen kernel of the Wilson-flow differential equation. It may still be possible to device a useful evolution equation able to distinguish between the confining and the conformal behavior of the theory in the infrared. 
 
\section{Topology and the $U(1)$ axial anomaly}
\label{sec:topology}
In order to identify all the relevant properties of the lower edge of the conformal window we need to determine  the role that each symmetry plays in this context. In particular, three symmetries or lack thereof and their order parameters should be considered: 
conformal symmetry, chiral (flavored) symmetry $SU(N_f)_L\times SU(N_f)_R$ and the $U(1)$ axial anomaly. Despite the fact that a true order parameter for confinement is absent in QCD, we can still identify confinement with the presence of a mass-gap, i.e, a glueball mass in quenched QCD. The lower edge of the conformal window  is then also the point where the properties of glueball operators change and the mass-gap disappears, as in Ref.~\citenum{Bochicchio:2013aha}. 

In this section we discuss the topology, the way the $U(1)$ axial anomaly manifests and its interplay with chiral  and conformal symmetry, in the conformal window and across its lower edge. 
Two quantities that are relevant to understand the fate of the $U(1)$ axial anomaly are the topological charge $Q$ 
\be
\label{eq:Qdef}
Q\,=\,\frac{1}{32\pi^2}\int_V\,d^4x\,G_{\mu\nu}^a\tilde{G}_{\mu\nu}^a\, ,
\ee
with $\tilde{G}$ the dual of the gauge-field strength tensor $G_{\mu\nu}$, and the topological susceptibility $\chi_t$, i.e., the second moment of the topological charge distribution
\be
\chi_t\,=\,\frac{\langle Q^2\rangle}{V}\,=\,-\frac{1}{V} \left. \left (\frac{1}{Z}\frac{\partial^2Z}{\partial\theta^2}\right )\right\vert_{\theta =0}
\ee
with the QCD partition function $Z$ dependent on the QCD $\theta$-angle.  
The $U(1)$ axial anomaly can be defined as the nonzero contribution to the divergence of the flavor singlet axial current
\be
\label{eq:axial}
\partial_\mu J_\mu^5\, = \,\frac{g^2N_f}{16\pi^2}\,G_{\mu\nu}^a\tilde{G}_{\mu\nu}^a \,+{\mbox{explicit fermion mass contribution}}.
\ee
This is an operator identity and it implies that the axial current is conserved, i.e., no anomaly  in two limits:
$N_f=0$ and $N=\infty$. The latter limit becomes evident in the Witten-Veneziano formula, where the anomaly contribution to the mass of the $\eta'$ meson only appears at $O(1/N)$. We add that results in the large-$N$ limit are directly  relevant in the formulation of AdS/CFT arguments. 

What is the fate of the $U(1)$ axial anomaly, what is the distribution of the topological charge $Q$, and what is the value of $\chi_t$ in the conformal window? To give what is maybe a straightforward answer, we first clarify how many relevant, in the renormalisation group sense, order parameters are at the lower edge of the conformal window.  
By dimensional arguments and the effective lagrangian description \cite{Pisarski:1983ms}  the $U(1)$ axial order parameter 
\be
\langle \det{\bar{\psi}^f_L\psi^f_R}\rangle\,\sim\, \Lambda^{3N_f}
\ee
becomes irrelevant for infrared physics for $N_f>2$, while the chiral symmetry order parameter $\langle\bar{\psi}_L\psi_R\rangle$ is always relevant. This already suggests that one single phase transition should be associated with the restoration of chiral symmetry and the effective restoration of the $U(1)$ axial symmetry\footnote{While the $U(1)$ axial anomaly in Eq.~(\ref{eq:axial}) is an operator identity and does not vanish in QCD for any finite $N_f$ and $N$, its effect may be absent in the 
 $n$-point functions of the theory, at least for $n=2$,  and one talks about the effective restoration of the symmetry.} for $N_f>2$.
This holds true for the chiral phase transition of finite temperature QCD with $N_f>2$, as well as  the conformal window with $N_f^c>2$. We can even add that the approximate statements that apply to the near-conformal system at $T>T_c$ become exact for the massless theory at the IRFP inside the conformal window.  

We now recall 
 the relation of the topological charge $Q$ to the eigenmodes and eigenvalues of the Dirac operator, the relation of the latter to the chiral condensate and the decomposition of the meson two-point functions in terms of the Dirac eigenvalues. 

The Atiyah-Singer index theorem \cite{Atiyah:1963zz} 
expresses the index of an elliptic operator of any compact oriented differentiable manifold in topological terms. 
While Eq.~(\ref{eq:Qdef}) is in terms of the gauge fields, the index theorem provides the topological charge in terms of the index of the Dirac operator, part of the QCD fermion sector: $Q$ can be written as the sum of the chiralities of the eigenmodes of the Dirac operator
\be
\label{eq:index}
Q\,=\, \sum_s\,\chi_s\,=\,n_+\,-\,n_-\, ,
\ee
where $n_\pm$ indicates the number of zero modes with chirality $+1$ and $-1$, respectively. In fact, 
since the zero modes of the (massless) theory have chirality $\chi_0=\pm 1$, while the nonzero modes have chirality $\chi_{(s\neq 0)} =0$, one should conclude that the topological charge is nonzero (and integer) only in the presence of zero modes. 
To summarize, the index theorem tells that any field configuration with $Q\neq 0$ has at least one eigenfunction of vanishing eigenvalue, and thus vanishing fermion determinant. 

The Banks-Casher relation \cite{Banks:1979yr} 
\be
\label{eq:BanksCasher}
\langle 0|\bar\psi\psi |0\rangle\, =\,-\lim_{m\to 0} m\,\int_{-\infty}^{\infty}\,d\lambda\,\frac{\rho (\lambda )}{\lambda^2+m^2}\,
\begin{cases}
\,=\, -\pi\,\rho (0), & \mbox{if } \rho(0)\neq 0 \\
\,\propto\, m^\alpha , &\mbox{if } \rho(\lambda)\propto\lambda^\alpha
\end{cases}
\ee
expresses the order parameter of chiral symmetry $\langle 0|\bar\psi\psi |0\rangle$ in terms of the Dirac eigenvalue distribution $\rho (\lambda )$. The limit $V\to\infty$ taken before the limit $m\to 0$ in the r.h.s. is mandatory in the presence of spontaneously broken chiral symmetry; it is not so if chiral symmetry is exact. Eq.~(\ref{eq:BanksCasher}) tells that the vanishing/appearance of the chiral condensate corresponds to the vanishing/appearance of $\rho (0)$, i.e., the Dirac eigenvalue level density at $\lambda =0$. 
Note that the Banks-Casher relation carries no explicit information on the $N_f$ dependence. 

The third useful ingredient is  the decomposition of the meson two-point functions in terms of Dirac eigenvalues; the analysis is based on Ref.~\citenum{Kogut:1998rh}, see also Ref.~\citenum{Deuzeman:2012ee}. The degeneracy pattern of the pseudoscalar  isovector ($\pi$),  scalar isosinglet ($\sigma$),  scalar isovector ($\delta$) and  pseudoscalar isosinglet ($\eta'$) two-point functions when $SU(N_f)_L\times SU(N_f)_R$ or $U_A(1)$ are restored
\[
\begin{array}{ccc}
C_{\vec{\pi}} &
\xleftrightarrow{\mbox{\normalsize $SU(N_f)~~$  }} &
C_\sigma \\[0.2cm]
U_A(1)\left\updownarrow\rule{0cm}{1.0cm}\right.
\phantom{U_A(1)}
&&
\phantom{U_A(1)}\left\updownarrow\rule{0cm}{1.0cm}\right.\phantom{U_A(1)}\\[0.2cm]
C_{\vec{\delta}} & \xleftrightarrow{\phantom{\mbox{\normalsize $SU(N_f)~~$  }} }&
C_{\eta'}
\end{array}
\]
is a useful vademecum. 
When chiral symmetry is exact, for  $T>T_c$ or in the conformal window, the ordering of the infinite volume and zero mass limits is irrelevant and the spectral decomposition of the meson correlators reads\cite{Kogut:1998rh}
\ba
\label{eq:CF}
C_{\vec{\pi}}&=&  -(Q =0) + (Q =\pm 1)  \nonumber\\
C_{\vec{\delta}}&=& - (Q =0) - (Q =\pm 1)  \nonumber\\
C_{\sigma}&=&  -(Q =0) - (Q =\pm 1) + {N_f} (Q =\pm 1)  \nonumber\\
C_{\eta'}&=&  -(Q =0) + (Q =\pm 1) - {N_f} (Q=\pm 1)\, , 
\ea
where the $Q=0,\,\pm 1$ terms are a shorthand notation for  the zero and $\pm 1$ topological charge sector contributions, and  only
the terms proportional to $N_f$ in $C_{\sigma}$ and $ C_{\eta'} $ come from disconnected contributions. 
 The case $N_f=2$ is special: exact chiral symmetry, i.e. the degeneracy of the chiral partners $C_{\vec{\pi}}=C_{\sigma}$ and $C_{\vec{\delta}}=C_{\eta'}$ is automatically verified, while  the degeneracy of the $U(1)$ axial partners, i.e.,  $C_{\vec{\pi}}=C_{\vec{\delta}}$ and $C_{\sigma}=C_{\eta'}$ requires the vanishing of the contributions from the nonzero topological charge sector. 
 For $N_f>2$, restored chiral symmetry itself requires the absence of $(Q=\pm 1)$ contributions  and it implies the degeneracy of the $U(1)$ axial partners. This is consistent with the fact that the chiral and $U(1)$ axial order parameters are both relevant for $N_f=2$, and their restoration can thus occur separately, while only the chiral order parameter is relevant for $N_f>2$. 
 One crucial difference between theories below the conformal window with $T>T_c$, and theories in the conformal window is that all contributions of irrelevant operators\footnote{It does not apply to dangerously irrelevant operators.} to $n$-point functions vanish {\em exactly} at the IRFP in the conformal window. 
 In other words, the $U(1)$ axial anomaly is not manifest at the IRFP. 
 
To understand the fate of $Q$ and $\chi_t$  we observe that the formulae valid for the symmetry-restoration region $V\Sigma m\ll 1$ of QCD with $N_f\geq 3$ degenerate flavors\footnote{The low-energy constant $\Sigma$ is related to the infinite volume fermion condensate in the chiral limit $\langle 0|\bar\psi\psi |0\rangle =-\Sigma$ at $\theta =0$.}  can to a certain extent be applied to the IRFP. 
In practice, we suggest that one can take the results in Ref.~\citenum{Leutwyler:1992yt} in the limit $V\Sigma m\rightarrow 0$\footnote{Given that the ordering of the infinite volume and zero mass limits does not matter in the restored phase, we can take $m\to 0$ first.}. 
The chiral condensate of QCD for $V\Sigma m\ll 1$ reads \cite{Leutwyler:1992yt}
\be
\label{eq:cond}
\langle \bar{\psi}_L\psi_R\rangle = -\frac{V\Sigma^2}{4N_f}\, m+\ldots ~~~(N_f\geq 3)\, ,
\ee
in contrast with $\langle \bar{\psi}_L\psi_R\rangle =-(\Sigma/2)e^{-i\theta}$ in the opposite limit $V\Sigma m\gg 1$.
The $\theta$-vacuum angle dependence has disappeared in Eq.~(\ref{eq:cond}), the condensate is proportional to the fermion mass\footnote{Note that we cannot obtain the power-law dependence with exponent $\delta$ of Eq.~(\ref{eq:EoS}) valid at a critical point.} and it decreases as $N_f$ increases, a screening effect. Also, the chiral condensate is exclusively due to nonzero modes in the trivial topology sector $Q=0$.

From   Ref.~\citenum{Leutwyler:1992yt} we also learn that
the  partition function at fixed winding number is 
$\propto  (V\Sigma m)^{|Q |N_f}$ for $V\Sigma m\ll 1$, while Gaussian for $V\Sigma m\gg 1$. 
 In both cases the topological susceptibility $\chi_t ={\Sigma m}/{N_f}$ to leading order in $m$, is proportional to the fermion mass and inversely proportional to $N_f$, consequence of the fact that  fermions drastically modify the gluodynamics by introducing a leading order fermion mass dependence in the topological quantities. 
Also,  the suppression of large $Q$ sectors increases with $N_f$, i.e., the shift towards larger eigenvalues increases with $N_f$. 

The interplay of gluodynamics and fermions is even more evident when taking the $N\to\infty$ limit with finite $N_f$ in mass-degenerate QCD\cite{Leutwyler:1992yt}. 
In this limit, fermion degrees of freedom strongly affect the properties of the partition function in the symmetry-restoration region  $V\Sigma m\ll 1$. On the contrary, outside this region the probability for finding field configurations of large winding number and the topological susceptibility are the same as in gluodynamics. Gluodynamics itself changes drastically for $T>T_c$. Lattice results\cite{DelDebbio:2004rw, Bonati:2013tt}  are consistent with an exponential suppression at finite $N$ of the Yang-Mills topological susceptibility $\chi_t\sim \exp{(-\gamma(T) N)}\cos{\theta}$; it vanishes in the large-$N$ limit\footnote{For $T<T_c$ the topological susceptibility of pure Yang-Mills remains nonzero at large-$N$ and it guarantees the presence of the anomaly contribution to the $\eta'$ mass, $m_{\eta'}^2 = O(1/N)$ via the Witten-Veneziano formula $f_\pi^2 m_{\eta'}^2 =2 N_f \chi_t$.}.  
This suppression is also consistent with the exponential suppression of instantons, which carry nonzero topological charge. Refs.~\citenum{Parnachev:2008fy, Zhitnitsky:2013wfa} have suggested that an analogous suppression occurs at the lower edge of the conformal window. 

Our discussion uses instead the fact that topology is drastically affected by fermions, that chiral symmetry is exact in the conformal window, that  the $U(1)$ axial order parameter is infrared irrelevant, the absence of zero modes of the Dirac operator, and that all VEVs of dimensionful quantities should vanish at the IRFP\footnote{This is true also for $\langle G\tilde G\rangle$. In QCD, instantons contribute to the $\theta$-vacuum expectation value as $\langle \theta |G\tilde G|\theta\rangle \propto K\,\sin{\theta} e^{-S_0}$, $S_0=8\pi^2/g^2$ and K contains the fermion determinant. This VEV vanishes in QCD because $K=0$ due to the presence of vanishing eigenvalues.}. 
All these properties suggest that the topological charge is zero at the IRFP, as well as the topological susceptibility. 
This means that there are no instantons (in contrast to instantons of all sizes in QCD). Cluster decomposition is satisfied, differently from what happens when taking a fixed topological charge sector in QCD\footnote{Zero topological charge in QCD can be made by a widely separated instanton with $+n$ and anti-instanton with $-n$ which would spoil cluster decomposition.}. 
In some sense, the same result is obtained in the infinite temperature limit $T\rightarrow\infty$ of QCD, where instantons are squeezed until none contributes and $U(1)$ axial symmetry becomes exact.    

A lattice formulation of the theory inside the conformal window will only approximately reproduce the properties so far discussed, and a numerical study will need to consider corrections to the physics of the IRFP induced by finite lattice spacing, finite volume and a non vanishing fermion mass. 
\section{The spectrum and conformal symmetry}
\label{sec:spectrum}
Understanding the breaking pattern of conformal symmetry  close to the Planck scale, i.e., at an energy scale where gravity is relevant and it is to be included in the complete lagrangian of the universe,  is likely to shed light on the way short distances relate to large distances within a unified description of physical forces.  
It is also an appealing possibility that answers are in large part to be found in the quantum field theoretical framework of which many features are familiar to us.  
 
In which manner is conformal symmetry broken? The spontaneous breaking of conformal symmetry is appealing in many respects, though we do not easily find examples of it in known physical systems. QCD in its present formulation is never conformal, it deviates from the free (conformal) theory in a highly nontrivial way, encoded in the logarithms of asymptotic freedom; in this example, as many others,  conformal symmetry is explicitly broken due to the presence of conformally non invariant terms in the renormalized lagrangian. 
The situation can be different once we couple matter fields to gravity; in the end conformal symmetry is a symmetry of spacetime itself. It is also plausible to think that we need to first formulate the complete theory in curved spacetime and eventually recover known physics in flat spacetime, with possible corrections. To what extent these corrections modify known low-energy theories is unknown, while they should not spoil the agreement of the Standard Model as presently formulated with established experimental results. 

In this context, one interesting question is: What are the signatures that allow to determine which way conformal symmetry is broken? We can choose among explicit breaking, the spontaneous breaking of a global symmetry and the spontaneous breaking of a local, i.e., gauge symmetry; we have examples of all three in nature. The spontaneous breaking of conformal symmetry can be implemented analogously to known examples, and it carries the imprint of dynamical scalar degree(s) of freedom\footnote{This scalar may carry some different properties as compared to ordinary matter scalar fields.}. If conformal symmetry is treated as a global symmetry, then a massless Goldstone boson, the dilaton would arise analogously to pions in QCD, the Goldstone bosons of global chiral symmetry. We could also promote the conformal symmetry group, or a subgroup of it, to a local gauge symmetry. This is a highly non trivial change, whose imprint would be a Brout-Englert-Higgs mechanism and the associated scalar boson. The idea of local gauge symmetry (local scale invariance) has been pursued recently.\cite{Hooft:2014daa, tHooft:2011aa}      

We can conceive that the scalar field of the spontaneously broken local or global conformal symmetry  will couple to matter fields of the complete theory for gravity, electroweak and strong interactions as dictated by the underlying symmetries, which now include conformal symmetry. 
Such scalar is therefore a dynamical degree of freedom which may leave its imprint at all scales. The breaking happens only once, in the complete theory, and one needs to uncover the renormalisation-group (RG) flow of all couplings from the Planck scale down to the electroweak symmetry breaking scale, and further down to low-energy QCD.  
It is not yet clear to which extent one should allow for additional explicit breakings at multiple energy scales below the Planck energy.  Certainly, all physical quantum anomalies responsible for explicit breakings should be preserved; an exception can be the conformal anomaly. 

This work focusses on the conformal window of QCD, and non-Abelian gauge theories in general, and it is legitimate to ask under which conditions is conformal symmetry lost at the lower edge of the conformal window and if a scalar degree of freedom related to its breaking may arise. 
Massless QCD in isolation, i.e. not embedded in the complete theory, seems to suggest a straightforward solution: the breaking of conformal symmetry  is in this case a consequence of the spontaneous breaking of chiral symmetry and the appearance of a mass-gap, i.e., confinement. The induced conformal symmetry breaking is thus explicit, no dilaton is needed; in other words, the vacuum expectation value of the trace of the energy-momentum tensor $T_\mu^\mu$ can become nonzero due to the contributions from the condensates $\langle GG\rangle$ and $\langle\bar{\psi}\psi\rangle$. 
This is not different in nature from the conformal symmetry breaking realized by the quantum corrections to classical chromodynamics. 
 
Perhaps physically more interesting is  the fate of the conformal window, once QCD is embedded in the complete theory.    
We have said that the scalar field(s) of the spontaneously broken global or local conformal symmetry at the Planck scale leaves its imprint in the dynamics, hence it affects the beta-functions of the complete theory.
The RG flow and fixed-point structure of the complete theory must be reconsidered in the presence of the new scalar degree(s) of freedom. A priori, the beta-function of the QCD gauge coupling itself $\beta (N, N_f)$ becomes a function of the scalar contributions, i.e., matter scalar fields and the dilaton. From the general construction in Ref.~\citenum{tHooft:2011aa} and the well known one-loop results\cite{Gross:1973ju} we deduce some relevant features in flat spacetime:
\begin{itemize}
\item[$\bullet$] The dilaton does not couple directly to $SU(N)$ Yang-Mills fields. It couples to matter scalar fields (the Higgs field and others) through the scalar potential and to fermions through a purely imaginary Yukawa term $(i m_f/\Lambda_{Pl})$$\bar{\psi}\eta (x)\psi$. This term disappears if fermions are massless, $m_f=0$. If  matter scalar fields are also absent, low-energy strong interactions do not talk to the dilaton. 
\item[$\bullet$] In the presence of scalar fields the gauge coupling beta-function $\beta (g)$ is modified; in particular, to find the fixed points one needs to consider the system of differential equations $\beta (g, \lambda_i, y_i)=0$ that includes all scalar fields self-couplings $\lambda_i$ and the Yukawa couplings $y_i$. 
\item[$\bullet$] Matter scalar fields usually screen gauge forces as fermions do. Thus, the coefficient of the one-loop beta-function $\beta (g) =-b_0 g^3$ becomes $b_0 =11/3 N -2/3 N_f -1/6 N_s$\cite{Gross:1973id, PhysRevLett.30.1346, Gross:1973ju}, where $N_s$ does {\em not} include the dilaton which does not couple directly to YM fields. One concludes that the loss of asymptotic freedom, $b_0=0$, is anticipated by matter scalars to a lower $N_f$ and the upper edge of the conformal window is shifted, unless asymptotic freedom has been fully destroyed by the scalar self-couplings $\lambda_i$.  
\item[$\bullet$] At two-loops in QCD $\beta(g)=-b_0 g^3 -b_1 g^5$ and an IRFP arises at $g_*^2 =-b_0/b_1$, as long as $b_0>0$ and $b_1<0$. A sign change of $b_1$ implies the disappearance of the IRFP and would  signal the lower edge of the conformal window $N_f^c$. Corrections to $b_1$ in the presence of matter scalar fields and the dilaton are more involved and depend on the details of the theory, its group and representation structure. 
We can still say that matter scalar fields will act analogously to fermions in screening gauge forces;  all in all the conformal window would be shifted to lower $N_f$ and its width changed, or disappear due to scalar self-couplings.  
\item[$\bullet$] 
The dilaton is special. Interestingly, its Yukawa coupling $y$ to fermions can contribute to $\beta (g)$ at $O(y^2g)$\cite{tHooft:2011aa} and it being imaginary, it carries a sign opposite to the other Yukawa contributions. On the other hand, if $y$ is proportional to the fermion mass its contribution vanishes if fermions are massless. \end{itemize}
To summarize, if fermion mass terms are turned on via, e.g.,  a Brout-Englert-Higgs mechanism, then the Yukawa coupling of the dilaton to fermions is turned on. It becomes then phenomenologically interesting to study small mass perturbations to a fixed point of the massless theory.   
  
\begin{figure}[tbp]
\begin{center}
\includegraphics[width=3in]{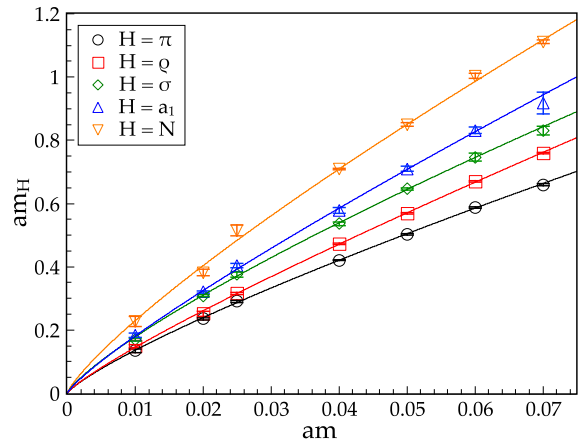}
\end{center}
\caption{Spectrum of the $N_f=12$ theory from Ref.~\citenum{Lombardo:2014pda}. }
\label{fig:allstates}
\end{figure}

\begin{figure}[tbp]
\begin{center}
\includegraphics[width=3in]{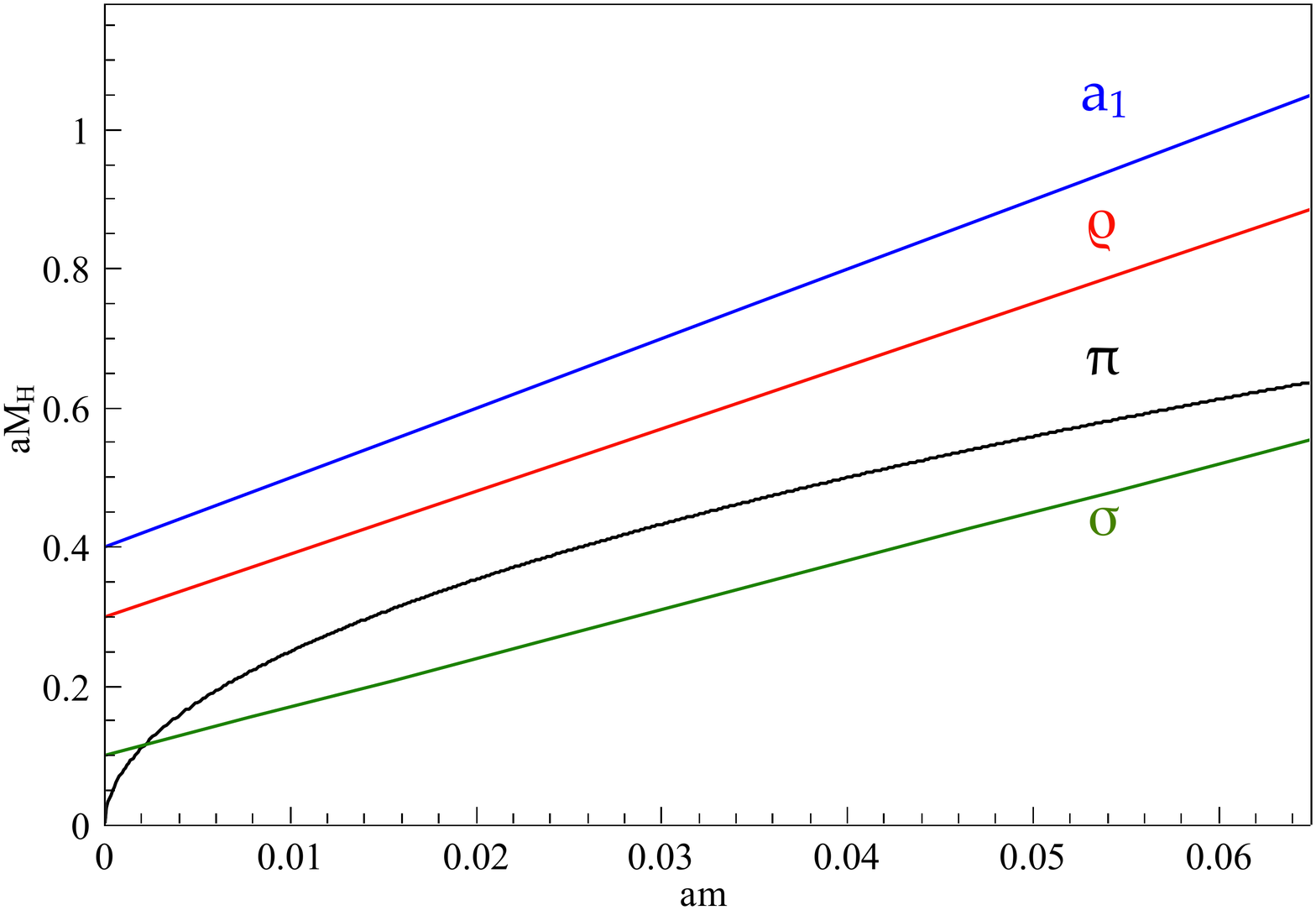}
\end{center}
\caption{Illustration of $\pi -\sigma$ inversion near the conformal window, away from the chiral limit.}
\label{fig:spectrum_below}
\end{figure}

Turning back to massless QCD in isolation, we comment on its spectrum when adding a small fermion mass perturbation and varying $N_f$ across $N_f^c$. 
QCD with $N_f=12$ massless flavors can be taken as a prototype of a theory inside the conformal window. Its  particle spectrum in the presence of a fermion mass perturbation  has been theoretically discussed in Ref.~\citenum{Lombardo:2014pda} with analogies to quantum critical phenomena. A perturbative mass deformation at the IRFP allows to parametrize violations of universal scaling and to determine on the lattice the anomalous dimension of the fermion mass operator at the IRFP. We find $\gamma_m\sim 0.25$ \cite{Lombardo:2014pda}; note that a value $\gamma_m\ll 1$ already  suggests that $N_f=12$ is not close to the lower edge of the conformal window, in agreement with  our recent results\cite{daSilva:2015vna} that place the endpoint between $N_f=8$ and $N_f=6$.   
The ordering of states, pseudoscalar (PS), vector (V), scalar (S), axial (PV) mesons and the nucleon (N) for twelve flavors does not lead to surprises, see Fig.~\ref{fig:allstates}.
All masses approach zero in the chiral limit with a universal power-law, corrected by mass-induced small violations of universal scaling \cite{Lombardo:2014pda}. This behavior signals exact chiral symmetry and the presence of a fixed point where conformal scaling holds. 
One obvious constraint below the conformal window, $N_f<N_f^c$, is that  the spectrum of the zero-temperature theory at a given $N_f$ cannot qualitatively differ from that of zero-temperature QCD, because they share the same underlying symmetries. Massless pseudoscalar mesons, the pions, are the Goldstone bosons of the spontaneously broken chiral symmetry, while the vector, axial and scalar mesons have a nonzero mass in the chiral limit. 
Away from the chiral limit, an ordering of states different from real world QCD can occur, provided all constraints implied by the spontaneously broken chiral symmetry and the $U(1)$ axial anomaly are satisfied. Some of these constraints, of immediate use in lattice studies, are i) the non degeneracy of chiral partners, which translates in the (pseudo)scalar sector into a constraint on the chiral cumulant $R_\pi\underset{m\to 0}{\rightarrow}0$, ii) the Gell-Mann-Oakes-Renner (GMOR) relation \cite{GellMann:1968rz}, iii) the non degeneracy of $U(1)$ axial partners. 
Generally, even an inversion of the pseudoscalar ($\pi$) and scalar ($\sigma$) states should be possible away from the chiral limit as shown in Fig.~\ref{fig:spectrum_below}, without violating any of the constraints. 
No dilaton will be present for QCD in isolation, while a scalar state numerically, but not parametrically, lighter can occur just below the conformal window. We add that recent AdS/CFT constructions \cite{Kutasov:2011fr} also fail to produce a parametrically lighter scalar. 
If a conformal phase transition occurs at the lower edge of the conformal window, the corresponding preconformal scaling would leave its imprint also in the spectrum, still ensuring that its properties  smoothly merge with those of the QCD spectrum\,---\,because no change of underlying symmetries separates the preconformal region and QCD.

\section{Conclusions}

Some observables change drastically across the lower edge of the conformal window of non-Abelian gauge theories. We have identified them and discussed their properties inside and below the conformal window; nonperturbative (lattice) studies of many of these properties would be desirable. Throughout, we have used the idea that the drastic change from QCD to theories inside the conformal window results from the interplay of chiral symmetry and confinement, and we discussed the somewhat different role of the $U(1)$ axial anomaly and conformal symmetry itself. 

Within the gauge sector, the scalar glueball operator exhibits  the nice feature that its anomalous dimension at the IRFP is constrained by the nonrenormalization of the energy-momentum tensor. The way this anomalous dimension varies along the IRFP line is a direct probe of the mechanism for the loss of conformal symmetry. 
Interestingly, perturbation theory and large-$N$ arguments predict a variation with $N_f$ opposite to the one associated with a UV-IR merging mechanism. 
In addition to the study of the two-point function of the scalar glueball operator in order to determine its anomalous 
dimension at the IRFP and its variation with $N_f$, 
we commented on the possibility that  a Wilson-flow equation with an appropriately devised  evolution kernel could discriminate between a conformal and a confining behavior in the infrared. 

Topology is a powerful probe of the conformal window. 
We have suggested a close analogy between high-temperature massless QCD with $N_f>2$, $T>T_c$ and a theory at the IRFP in the conformal window. One crucial difference between the two is  the actual existence in the latter of an IRFP, which makes approximate statements exact.
It is also observed that the commutativity of the infinite volume and zero mass limits in the chirally symmetric phase allows to a certain extent to establish an analogy with the symmetry-restoration region of QCD, where one works in the limit $V\Sigma m\ll 1$.   
The restoration of chiral symmetry, the infrared irrelevance of the $U(1)$ axial order parameter for $N_f>2$,  the exact  disappearance of irrelevant operators at the IRFP, the absence of zero eigenmodes of the Dirac operator and the vanishing of all vacuum expectation values suggest consistently that the topological charge and the topological susceptibility vanish at the IRFP; in other words, the effect of the anomaly is irrelevant for infrared physics which is all contained in the trivial topology sector. Fermions provide the dominant effect inside the conformal window, canceling all effects from gluodynamics. 
On a lattice, mass corrections jointly to lattice spacing effects and finite spacetime volume  force the system to be never exactly at the IRFP, so that topological quantities will not be exactly zero, though drastically suppressed as compared to QCD and theories just below the conformal window. 

If conformal symmetry is spontaneously broken at the energy scale where gravity matters, then its associated scalar degree of freedom plays a role in the complete theory at all scales. For QCD in isolation it is plausible that conformal symmetry is explicitly broken as a consequence of the spontaneous breaking of chiral symmetry and confinement, and no dilaton arises. The spectrum just below the conformal window allows for inversion in the ordering of states away from the chiral limit, but it does not allow for a parametrically lighter scalar meson due to the constraints imposed by the spontaneously broken chiral symmetry, i.e., the vanishing of the chiral cumulant in the massless limit due to the pseudoscalar Goldstone bosons of chiral symmetry. 
It remains  interesting and challenging to consider QCD as well as the complete Standard Model for strong and electroweak interactions embedded in a hypothesized complete theory that couples them to gravity, and study which way a spontaneously broken conformal symmetry modifies low-energy interactions. 
\section*{Acknowledgments}
I thank the organizers of SCGT15 for the interesting workshop and warm hospitality and my collaborators A. Deuzeman, M.P. Lombardo, K. Miura, T. Nunes da Silva and L. Robroek for all the inspiring discussions and work together. 
 
\bibliographystyle{ws-procs975x65}
\bibliography{references}

\end{document}